\documentclass[11pt]{article}

\usepackage{graphicx}% Include figure files
\usepackage{epsfig}
\usepackage{amssymb}

\begin{document}

\title{\bf The Structure of Electroweak Corrections\\ Due to Extended Gauge Symmetries}

\author{R. Sekhar Chivukula$^{1\,*}$, Hong-Jian He$^2$, \\
Joseph Howard$^1$, and Elizabeth H. Simmons$^1$\footnote{Address after 8-16-03: Department of Physics and Astronomy, Biomedical Physical Sciences Building, East Lansing, MI, 48825-2320}\ \thanks{e-mail addresses: sekhar@msu.edu, hjhe@physics.utexas.edu, howard@bu.edu, esimmons@msu.edu}\\ \\
$^1$ Physics Department, Boston University \\
590 Commonwealth Ave., Boston, MA 02215 USA\\ \\
$^2$ Center for Particle Physics and Department of Physics\\ 
University of Texas at Austin, Austin, TX 78712, USA}

\date{\today}

\maketitle

\bigskip
\begin{picture}(0,0)(0,0)
\put(355,295){BUHEP-03-16}
\put(355,280){UTHEP-03-10}
\end{picture}

%%%%%%%%%%%%%%%%%%%%%%%%%%%%%%%%%%%%%%%%%%%%%%%%%%%%%%%%%%%%%%
%%%%%%%%%%%%%%%%%%%%%%%%%%%%%%%%%%%%%%%%%%%%%%%%%%%%%%%%%%%%%%

\begin{abstract}

\end{abstract}
This paper studies models with extended electroweak gauge sectors of the form $SU(2) \times SU(2) \times U(1) \times [SU(2){\ \rm or}\ U(1)] $.  We establish the general behavior of corrections to precision electroweak observables in this class of theories and connect our results to previous work on specific models whose electroweak sectors are special cases of our extended group.

\newpage

\section{Introduction}

Understanding the dynamics of mass generation requires looking
beyond the standard model of particle physics.
Models with extended electroweak gauge sectors play a large 
role in recent work on the origins of electroweak symmetry breaking, 
the large top quark mass and the top-bottom mass splitting \cite{Hill:2002ap,Simmons:2002zi}.
Some theories include an extended $SU(2)\times SU(2)$ structure for the
weak interactions; the fermions may transform under the 
groups in a variety of flavor non-universal ways.   Examples include
non-commuting extended technicolor (NCETC) \cite{Chivukula:1994mn,Chivukula:1995gu}, topflavor \cite{Muller:1996dj,Malkawi:1996fs,He:1999vp}, and the 
ununified standard model (UUM) \cite{Georgi:1989ic,Georgi:1989xz} and
related theories \cite{Li:2002ps}.
Other theories include an extended $U(1)\times U(1)$ structure.  Examples include 
topcolor-assisted technicolor \cite{Hill:1991at,Hill:1994hp}, 
flavor-universal topcolor-assisted-technicolor \cite{Lane:1998qi,Popovic:1998vb}, topflavor-seesaw \cite{He:1999vp}, 
and string theories \cite{Cvetic:1997wu} or string-inspired models \cite{Cleaver:1998gc}.

This paper studies the phenomenology of models with replicated
electroweak gauge symmetries, based on a framework with the
gauge structure $SU(2) \times SU(2) \times U(1) \times [SU(2){\ \rm or}\ U(1)] $.
We discuss flavor-universal theories of this type in detail and show how
to extend our results in a straightforward manner to flavor non-universal theories.

We begin, in section 2, by establishing the conventions we shall use and
demonstrate that many specific models discussed in the literature are
special cases of our general framework.  In section 3, we examine the
corrections the extended gauge sector causes in the low-energy current-current
interactions of both the charged and neutral gauge sectors.  We find that the
form of the corrections depends neatly on the charge assignments of the
fermions under the extended electroweak group.  In section 4, we perform
a similar analysis of the masses and couplings of the $W$ and $Z$ bosons
in our general framework.  Again, the fermion transformation properties
dictate the form of the leading corrections to the $Zff$ and $Wff$ couplings.
In section 5, we perform a global fit of precision electroweak data to the
theories in which the leading predicted corrections to observables are 
smallest.  We discuss our conclusions and their application to other models
in the literature in section 6.

\section{Replicated Electroweak Gauge Symmetries}

The general framework \cite{Chivukula:2003ea} for the models we will discuss is based on
the gauge structure $SU(2) \times SU(2) \times U(1) \times [SU(2){\ \rm or}\ U(1)] $, and is
 illustrated in the figure below.   This figure
employs the conventions of  ``moose''  notation \cite{Georgi:1985hf},
\begin{equation}
\lower20pt\hbox{\includegraphics[width=8cm]{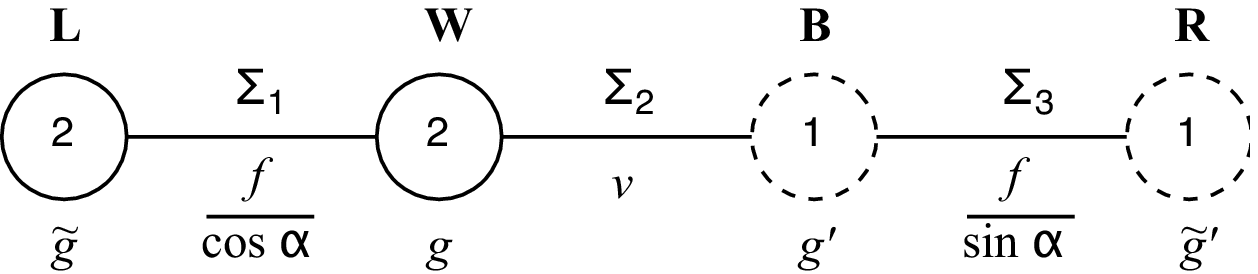}}
\end{equation}
in which the circles represent gauge
groups with the specified gauge coupling, and the solid lines
represent separate $(SU(2) \times SU(2)/SU(2))$ nonlinear sigma model
fields which break the gauged or global symmetries to which they are
attached. Each solid circle represents an $SU(2)$ or $U(1)$ group for which the entire global symmetry has been gauged; the dotted circle represents  
a global SU(2) in which only a U(1) subgroup has been gauged.

For convenience, the coupling constants of the gauge theories will
be specified by
\begin{equation}
\tilde{g}^\prime = {e\over \cos\theta\, \sin\phi}\,,\ \ 
g^\prime = {e\over \cos\theta\, \cos\phi}\,,\ \ 
g= {e\over \sin\theta\, \cos\omega}\,,\ \
\tilde{g}={e\over \sin\theta\, \sin\omega}\,,
\end{equation}
and the $f$-constants (the analogs of $f_\pi$ in QCD) of the nonlinear sigma
models by
\begin{equation}
{f\over \sin\alpha}\,,\ \ v\,,\ \ {f\over \cos\alpha}~.
\end{equation}

The nonlinear sigma fields will break the extended gauge group first (at scale $f$) to an intermediate electroweak group $SU(2) \times U(1)$ and then (at scale $v$) to the unbroken electromagnetic group $U(1)_{EM}$.  Hence the low-energy spectrum of the theory includes a massless photon, a set of $W$ and $Z$ bosons which closely resemble those of the standard model, and an extra set of heavy $W$ and $Z$ bosons corresponding to the  generators broken at scale $f$.

As we will see, the Lagrangian parameters $e$, $\theta$, and $v$, will
be approximately equal to the electric charge, weak mixing angle, and
Higgs expectation value in the one-doublet standard model. The scale
$f$ sets the masses of the extra gauge bosons, and the theory reduces
to the standard model in the limit $f\to\infty$.     The angle
$\alpha$ allows us to independently vary the breaking scale of the
duplicated $SU(2)$ or $U(1)$ gauge symmetries.  The angles
$\phi$ and $\omega$ determine the couplings of the gauge bosons which
become massive at scale $f$.  This analysis will assume that the ratio $\frac{v}{f} \ll 1$ since precision electroweak data already tells us that the observed $W$ and $Z$ bosons closely resemble those of the standard model.  

The symmetry structure of this model is similar to that proposed in the BESS
({\it Breaking Electroweak Symmetry Strongly})
model \cite{Casalbuoni:1985kq,Casalbuoni:1995qt}, an effective Lagrangian
description motivated by strong electroweak symmetry breaking.  The BESS model is
in turn an application of ``hidden local symmetry" to electroweak
physics \cite{Bando:1987br}.   Accordingly, we refer to our paradigm as
``Generalized BESS"  (GBESS).  The symmetry structure in the limit $f\to\infty$ is
precisely that expected in a ``technicolor''
model \cite{Weinberg:gm,Weinberg:bn}, and the theory has a custodial symmetry
in the limit where $g'$ and $\tilde{g}'$ go to zero.

GBESS is the simplest model of an extended electroweak
gauge symmetry incorporating both replicated $SU(2)$ and $U(1)$ gauge
groups. As such the electroweak sector of a number of models in the literature 
can be thought of as forming special cases of GBESS.   For example
{\it Noncommuting Extended Technicolor (NCETC)} \cite{Chivukula:1994mn,Chivukula:1995gu},  can be framed in the moose notation as follows:
\begin{equation}
\lower20pt\hbox{\includegraphics[width=8cm]{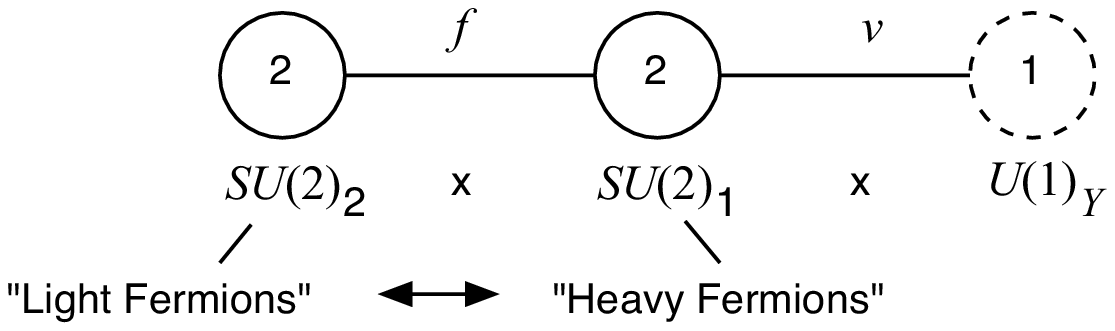}}
\label{eq:ncetcfig}
\end{equation}
where the gauge groups correspond, from left to right, to the $SU(2)_2 \times SU(2)_1 \times U(1)$ electroweak structure of the model.  In the ``heavy" case of NCETC, the fermions transforming under $SU(2)_2$ are those of the first and second generations as shown, while the third generation transform under $SU(2)_1$; in the ``light" case, the fermion assignments are reversed (per the double-headed arrow).  The electroweak gauge structure of the Ununified Standard Model \cite{Georgi:1989ic,Georgi:1989xz}  is the same,
\begin{equation}
\lower20pt\hbox{\includegraphics[width=8cm]{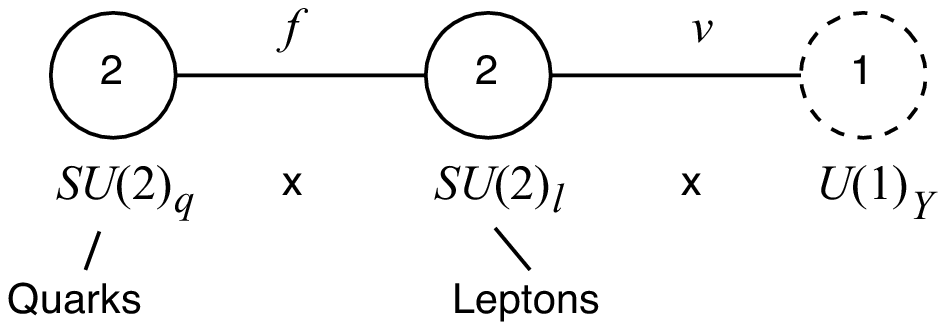}}
\end{equation}
but it is the quarks and leptons which now transform under separate $SU(2)$ groups, reflecting the distinct weak interactions of these species in this model.   As another example, the electroweak structure of 
 {\it topcolor} \cite{Hill:1991at,Hill:1994hp},  has the form
\begin{equation}
\lower20pt\hbox{\includegraphics[width=8cm]{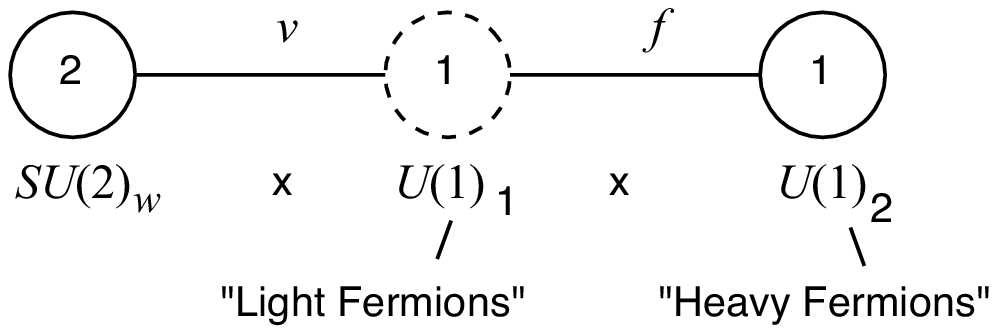}}
\end{equation}
which illustrates the separate hypercharge interactions of the 3rd generation and light fermions.  

Finally,   {\it electroweak $SU(3)$} \cite{Pisano:ee,Frampton:1992wt,Dimopoulos:2002mv} has the gauge structure $SU(3)\times SU(2)\times U(1)$ which can be represented as follows
\begin{equation}
\lower20pt\hbox{\includegraphics[width=7cm]{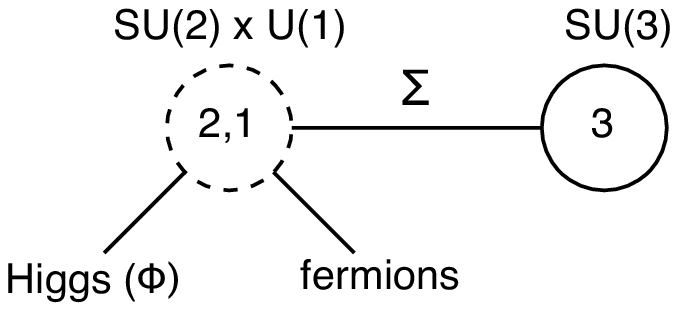}}
\end{equation}
where the dashed circle reflects the fact that the gauged $SU(2)\times U(1)$ is a subgroup of a global $SU(3)$ group.  Note that all fermions and the conventional Higgs field ($\phi$) are charged under this $SU(2)\times U(1)$.  The $\Sigma$ transforms as a $(2,3)_{-{1\over 2}}$ of the gauge symmetries.  If we focus on the $SU(2)\times U(1)$ subgroup of the $SU(3)$ gauge group which is preserved by the $SU(3)\times SU(3) / SU(3)$ chiral symmetry breaking structure, 
this model can be redrawn.
\begin{equation}
\lower20pt\hbox{\includegraphics[width=6cm]{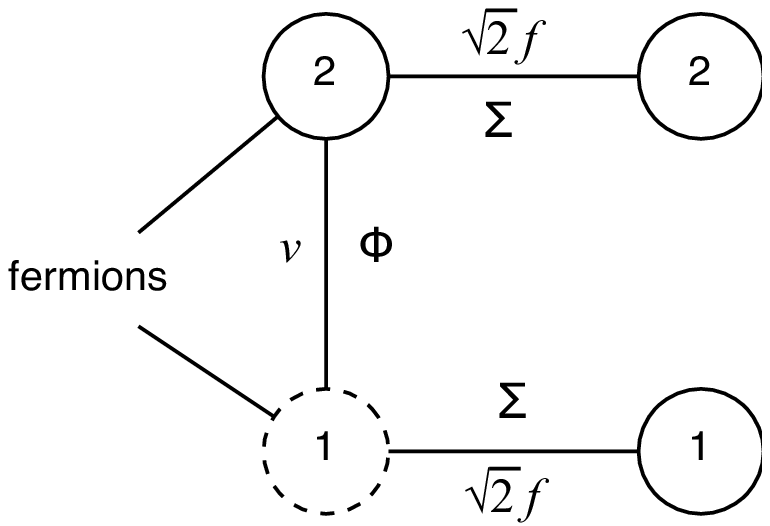}}
\end{equation}
This unfolds to reveal the linear moose form of GBESS once more, with the coupling values shown below
\begin{equation}
\lower20pt\hbox{\includegraphics[width=8cm]{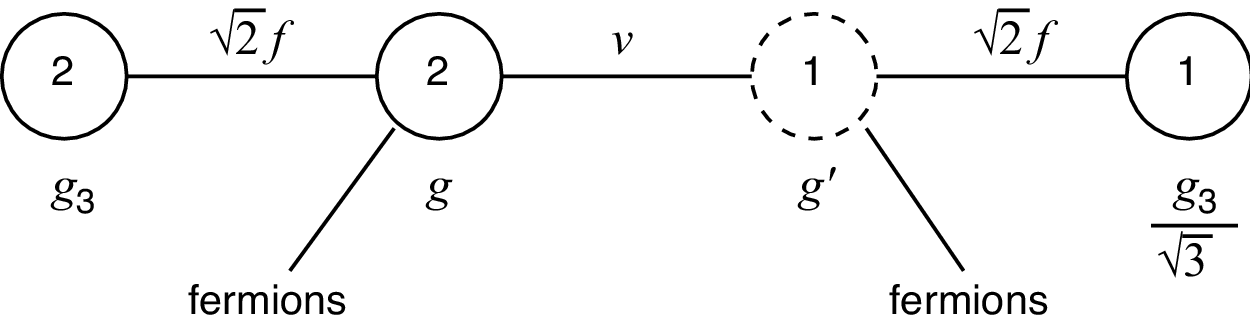}}
\end{equation}
and with the standard model fermions transforming under the inner two groups.
Note that for $g,g' \gg g_3$ one finds $\sin^2(\theta_W) \approx \frac{1}{4}$.
The general properties of precision electroweak constraints on these 
models\cite{Chivukula:1995gu,Chivukula:1996cc,Csaki:2002bz} can correspondingly be viewed as special cases of what follows, as we will discuss below.

\section{Low-Energy Interactions}

Important constraints on models with extended electroweak symmetries arise
from low-energy measurements, of which the most sensitive are the muon lifetime (used to determine $G_F$), atomic parity violation (APV),
and neutrino-nucleon scattering. In the usual fashion, we may summarize
the low-energy interactions in terms of four-fermion operators. The form
of these interactions will depend, however, on the fermion charge
assignments. For simplicity, we focus on 
models in which the fermion charge assignments are flavor
universal. To illustrate the model-dependence of the results,
we consider two examples.

First, we consider the case (hereafter, Case I) in which the ordinary fermions are charged
only under the two groups at the middle of the moose
\begin{equation}
\lower25pt\hbox{\includegraphics[width=8cm]{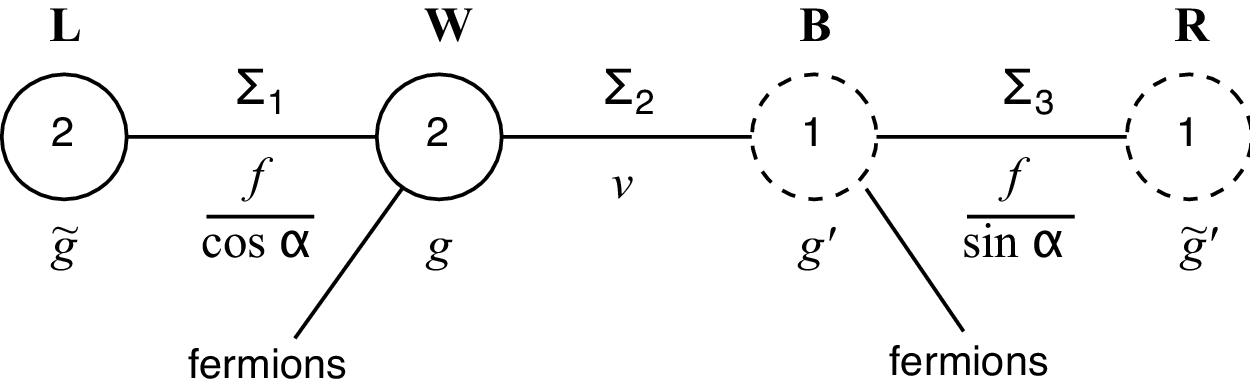}}\,.
\end{equation}
In this
case the charged current interactions may be computed to be
\begin{equation}
{L}^{CC} = -\,{2\over v^2}\, J^{\mu +} J^-_\mu\,,
\end{equation}
and the neutral current interactions to be 
\begin{eqnarray}
{L}^{NC} = &-&\,{2\over v^2}(J^\mu_3 - Q^\mu\, \sin^2\theta)^2  \nonumber \\ 
&-&\,{2 \over f^2}\,(\cos^2\alpha\,\sin^4\theta\,\sin^4\omega + \sin^2\alpha\,\cos^4\theta\,\sin^4\phi)\,Q^\mu Q_\mu~.
\end{eqnarray}
In these expressions, the currents $J^\mu_{\pm,3}$ and $Q^\mu$
are the conventional weak and electromagnetic currents.
From these, we see that the strength of $G_F$, APV, and neutrino
scattering is determined by $v$ in the usual way. Furthermore, comparing
the two equations, we see that the strength of the charged and
neutral current interactions, the so-called low-energy $\rho$
parameter, is precisely one (at tree-level). This last fact
is a direct consequence of the Georgi-Weinberg neutral current
theorem \cite{Georgi:1977wk}.

As an alternative, consider the case (hereafter, Case II) in which the $SU(2)$ charges
of the ordinary fermions arise from transforming under the gauge
group at the end of the moose ($SU(2)_L$ with coupling $\tilde{g}$)
\begin{equation}
\lower25pt\hbox{\includegraphics[width=8cm]{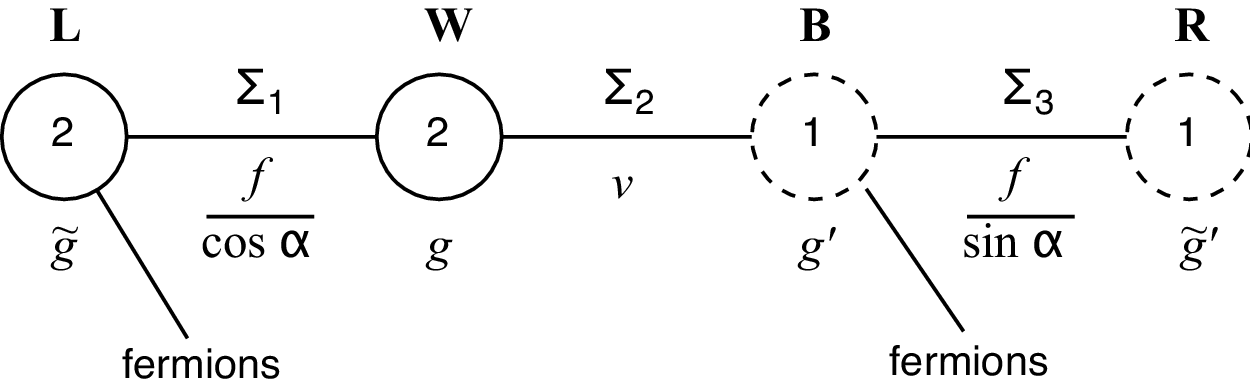}}\,.
\end{equation}
A calculation of the charged current
interactions yields
\begin{equation}
L^{CC}=-2\left({1\over v^2}+{\cos^2\alpha\over f^2}\right) J^{\mu +}J^-_\mu\,,
\end{equation}
while the neutral current interactions are summarized by
\begin{eqnarray}
{L}^{NC} = &-&\,{2\over v^2}(J^\mu_3 - Q^\mu\, \sin^2\theta)^2
 -\,{2\cos^2\alpha \over f^2}\,(J^\mu_3-\sin^2\theta\sin^2\omega\,Q^\mu)^2
\nonumber \\ &-&\,
 {2\sin^2\alpha\over f^2}\,(\cos^4\theta\, \sin^4\phi)\, Q^\mu Q_\mu~.
\end{eqnarray}
Several points in this expression are of particular note.  First, the value of $G_F$
as inferred from muon decay is no longer simply proportional to $v^{-2}$, but is instead proportional to $(v^{-2} + \cos^2\alpha f^{-2})$.  As
we shall see in the next section, this ultimately will give rise
to corrections to electroweak observables of order $(v/f)^2$
which are not suppressed by any ratios of coupling constants. Second,
in contrast to Case I, the strength of low-energy charged-
and neutral-current interactions are no longer the same.  
The strengths of the $J_3 J_3$
and $J^+ J^-$ portions of the interactions are still equal; this 
reflects the approximate custodial symmetry of the underlying model.  
The ratio of the strengths of the $J_3 Q$ and  $J^+ J^-$ terms, however, is
no longer simply proportional to $\sin^2\theta$, but depends also on the
angle $\omega$ relating the gauge couplings $g$ and $\tilde{g}$.

One can likewise define models in which the fermions couple to the group
at the right-hand end of the moose instead of to the gauged $U(1)$ group on
the interior.  Whether the global $SU(2)_R$ is fully gauged or only a $U(1)$ subgroup is gauged, the effect on the neutral-current weak interactions will be the same; the charged weak bosons of a gauged $SU(2)_R$ would not mix with those of $SU(2)_W$ and $SU(2)_L$.
We will refer to the situation where fermions are charged under groups W and R as Case III and
that where fermions are charged under groups L and R as Case IV.  The low-energy charged currents of Case III (Case IV) are identical to those of Case I (Case II).   The low-energy neutral currents may be written in the general form:
\begin{equation}
{L}^{NC} = -\,{2\over v^2}(J^\mu_3 - Q^\mu\, \sin^2\theta)^2
 -\,{2\cos^2\alpha \over f^2}\,(t_3 J^\mu_3- t_Q Q^\mu)^2
 -\,{2\sin^2\alpha \over f^2}\,(u_3 J^\mu_3- u_Q Q^\mu)^2
\label{eq:lowEnc}
\end{equation}
where the coefficients $t_3,\, t_Q,\, u_3,\, u_Q$ are summarized in Table \ref{tab-lowEnc}.  Note that in Cases III and IV the strenghts of the $J_3 J_3$
and $J^+ J^-$ portions of the interactions are no longer equal. 

\begin{table}[htb]
\begin{center}
\begin{tabular}{|l|c|c|c|c|}
\hline
Case & $t_3$ & $t_Q$ & $u_3$ & $u_Q$ \\
\hline\hline
I \ (W, B) &$0 $ & $s^2_\theta s^2_\omega $ & $0 $ & $ c^2_\theta s^2_\phi$ \\
II \ (L, B) &$1 $ & $s^2_\theta s^2_\omega $ & $ 0 $ & $ c^2_\theta s^2_\phi$ \\
III \ (W, R)&$ 0$ & $ s^2_\theta s^2_\omega $ & $ 1$ & $1 - c^2_\theta s^2_\phi $ \\
IV\ (L, R) &$ 1$ & $s^2_\theta s^2_\omega $ & $1 $ & $ 1 - c^2_\theta s^2_\phi $ \\
\hline\hline
\end{tabular}
\caption{Coefficients of the order $v^2/f^2$ corrections to low-energy neutral currents
(per equation \protect\ref{eq:lowEnc}) in Cases I-IV.  For each Case we note the gauge
groups ($SU(2)_{L\, {\rm or}\, W}$, $U(1)_{B\, {\rm or}\, R}$) under which the fermions are charged.  For brevity,
we abbreviate $\cos^2\alpha$ as $c^2_\alpha$, and $\sin^2\alpha$ as $s^2_\alpha$, and so forth.}
\label{tab-lowEnc}
\end{center}
\end{table}

\section{Z-Pole Constraints - General Structure}

Many of the most significant constraints on physics beyond the
standard model arise from precise measurements at the Z-pole.  To
interpret these measurements, we must compute the masses of the light $W$ and $Z$
bosons (which are the same in Cases I-IV) and their shifted couplings to ordinary fermions in terms of the
Lagrangian parameters. 

The charged weak bosons of the $SU(2)_L$ and $SU(2)_W$ groups of GBESS mix with one another\footnote{If one were to gauge the full global $SU(2)_R$ of GBESS, the charged weak
bosons of group $R$ would not mix with those of groups  $L$ and $W$ and would
simply have masses $M^2_{R^\pm}= {e^2 f^2\over 4\cos^2\theta}\left(\frac{1}{ \sin^2\alpha \sin^2\phi}\right)$.}   .    Diagonalizing the mass-squared matrix
\begin{equation}
\frac{e^2}{4\sin^2\theta}\,\frac{f^2}{\cos^2\alpha\,\sin^2\omega} \left( \begin{array}{cc} 1 & - \tan\omega \\ - \tan\omega & \tan^2\omega \left[1 + \frac{v^2\cos^2\alpha}{f^2}\right] \end{array} \right)
\end{equation}
for the charged $L$ and $W$  bosons 
yields masses for the light $W$ 
\begin{equation}
M^2_W = {e^2 v^2\over 4\sin^2\theta}
\left(1-\cos^2\alpha \sin^4\omega\,{v^2\over f^2}\right)
+ O\left({v^4\over f^4}\right)~,
\end{equation}
and its heavy counterpart (at leading order)
\begin{equation}
M^2_{L^\pm}= {e^2 f^2\over 4\sin^2\theta}
\left(\frac{1}{\cos^2\alpha\sin^2\omega\cos^2\omega} \right)
\end{equation}
 
All four neutral bosons in GBESS mix with one another.  Diagonalizing the mass-squared matrix  
(from which an overall factor of $[e^2 f^2 / 4 \cos^2\theta\sin^2\theta]$ has been removed)
\begin{equation}
\left( \begin{array}{cccc}
\frac{\cos^2\theta}{\cos^2\alpha\,\sin^2\omega} &  
-\frac{\cos^2\theta}{\cos^2\alpha\,\sin\omega\,\cos\omega}& 0&0\\[3mm]
-\frac{\cos^2\theta}{\cos^2\alpha\,\sin\omega\,\cos\omega}&
\frac{\cos^2\theta}{\cos^2\omega}\left[\frac{v^2}{f^2} + \frac{1}{\cos^2\alpha}\right] &
- \frac{v^2}{f^2} \frac{\cos\theta\,\sin\theta}{\cos\omega\,\cos\phi}& 0 \\[3mm]
0& - \frac{v^2}{f^2} \frac{\cos\theta\,\sin\theta}{\cos\omega\,\cos\phi} &
\frac{\sin^2\theta}{\cos^2\phi}\left[\frac{v^2}{f^2} + \frac{1}{\sin^2\alpha}\right] &
-\frac{\sin^2\theta}{\sin^2\alpha\,\sin\phi\,\cos\phi}\\[3mm]
0&0&-\frac{\sin^2\theta}{\sin^2\alpha\,\sin\phi\,\cos\phi}& 
\frac{\sin^2\theta}{\sin^2\alpha\,\sin^2\phi}  \end{array} \right)
\end{equation}
yields  a massless photon, a light $Z$ state with mass
\begin{equation}
M^2_Z = {e^2 v^2\over 4\sin^2\theta \cos^2\theta}
\left(1-(\cos^2\alpha \sin^4\omega+\sin^2\alpha\sin^4\phi)\,{v^2\over f^2}\right)
+\ldots
\end{equation}
and two heavy $Z$ states with leading-order masses
\begin{equation}
M^2_{N1}= {e^2 f^2\over 4\cos^2\theta}
\left(\frac{1}{\sin^2\alpha\sin^2\phi\cos^2\phi}\right)\end{equation}
\begin{equation}
M^2_{N2}= {e^2 f^2\over 4\sin^2\theta}
\left(\frac{1}{\cos^2\alpha\sin^2\omega\cos^2\omega}\right)
\end{equation}

By comparing our expression for $M^2_Z$ with our previous results 
for the relationships between $G_F$ and $v$ in Cases I and II, we can 
see that the fermion charge assignments have a major effect on the 
predicted values of electroweak observables.  The leading 
corrections to the standard model relation between $G_F$,
$\alpha$, $M^2_Z$, and the appropriately defined weak mixing angle
$\sin^2\theta_W$ are generically of order $v^2/f^2$ in Case II.  
In Case I, the dominant corrections are further suppressed by 
factors of  $\sin^4 \omega$ or $\sin^4\phi$.   Then, if one views 
the electroweak predictions of GBESS in terms of
{\it corrections} to the corresponding standard model results, these
corrections in Case I are (potentially)
suppressed by ratios of coupling constants relative to the size of
corrections in Case II. This leads generically to
weaker constraints in Case I models, meaning that new phenomena associated with 
such models (such as extra gauge bosons) can exist at lower energy scales.

In order to make predictions for electroweak observables,
we need to compute the couplings of the ordinary fermions to the light
gauge boson eigenstates. For Case I theories, we find
that the couplings to the left-handed fermions  to the $W$ are
\begin{equation}
{e\over \sin\theta}\left(1-\cos^2\alpha\,\sin^4\omega\,{v^2\over f^2}\right)+\ldots
\end{equation}
and for the $Z$ we find the couplings
\begin{eqnarray}
{e\over \sin\theta\cos\theta} 
\left[ 1-  ( \cos^2\alpha\,\sin^4\omega + \sin^2\alpha\,\sin^4\phi)\,
{v^2\over f^2} + \ldots  \right]\,T_3 \nonumber\\
 -\,{e\over \sin\theta\cos\theta} \left( \sin^2\theta-  \sin^2\alpha\,\sin^4\phi\,{v^2\over f^2} 
 + \ldots  \right)\,Q~.
\end{eqnarray}
Examining our expressions for the shifted gauge-boson 
masses and couplings,  we see that, for Case I, all corrections to standard
model predictions may be expressed in terms of two combinations
of Lagrangian parameters:
\begin{equation}
c_1=\cos^2\alpha\,\sin^4\omega\,{v^2\over f^2}\,,\ \ \ \ \ 
c_2= \sin^2\alpha\,\sin^4\phi\,{v^2\over f^2}~.
\label{cdefs}
\end{equation}
This allows us to compute bounds on model parameters in terms of fits
to $c_1$ and $c_2$, greatly simplifying the calculations.

For Cases II-IV, the leading corrections to gauge-boson couplings at order $v^2/f^2$ 
are suppressed by only two powers of $\sin\omega$ or $\sin\phi$.
To facilitate comparison of the forms of the couplings in the different
Cases, we write the general formula for $W$ couplings to left-handed fermions
\begin{equation}
{e\over \sin\theta}\left(1 + a_\pm {v^2\over f^2}\right)
\label{eq:gencoupw}
\end{equation}
and for $Z$ couplings to fermions
\begin{equation}
{e\over \sin\theta\cos\theta} 
\left[ 1 +  a_3 {v^2\over f^2}\right]\,T_3
 -\,{e\over \sin\theta\cos\theta} \left[ \sin^2\theta +   a_Q {v^2\over f^2}
\right]\,Q~.
\label{eq:gencoupz}
\end{equation}
and tabulate the factors $a_\pm,\, a_3,\,$ and $a_Q$ in Table \ref{tab-couplings}.
The general pattern is that when the fermions couple to a group on the interior (exterior) of the moose, the leading correction at order $v^2/f^2$ goes as $\sin^4$ ($\sin^2$) of the associated mixing angle.  Thus the corrections in models where fermions couple to an exterior group can be significantly larger than those in Case I.

\begin{table}[htb]
\begin{center}
\begin{tabular}{|l|c|c|c|}
\hline
Case & $a_\pm$ & $a_3$ & $a_Q$ \\
\hline\hline
I (W, B)&$ - c^2_\alpha s^4_\omega$ & $ - c^2_\alpha s^4_\omega -s^2_\alpha s^4_\phi$ & $- s^2_\alpha s^4_\phi$ \\
II (L, B)& $c^2_\alpha s^2_\omega c^2_\omega$ & $ c^2_\alpha s^2_\omega c^2_\omega- s^2_\alpha s^4_\phi $ & $ - s^2_\alpha s^4_\phi $ \\
III (W, R)& $ - c^2_\alpha s^4_\omega$  & $- c^2_\alpha s^4_\omega + s^2_\alpha s^2_\phi c^2_\phi $ & $ s^2_\alpha s^2_\phi c^2_\phi$ \\
IV (L, R)& $c^2_\alpha s^2_\omega c^2_\omega$ & $ c^2_\alpha s^2_\omega c^2_\omega + s^2_\alpha s^2_\phi c^2_\phi $ & $ s^2_\alpha s^2_\phi c^2_\phi$ \\
\hline\hline
\end{tabular}
\caption{Coefficients of the order $v^2/f^2$ corrections to fermion couplings to 
$W$ and $Z$ bosons (per equations \protect\ref{eq:gencoupw}, \protect\ref{eq:gencoupz}) in Cases I-IV.   For each Case we note the gauge
groups ($SU(2)_{L\, {\rm or}\, W}$, $U(1)_{B\, {\rm or}\, R}$) under which the fermions are charged.   For brevity,
we abbreviate $\cos^2\alpha$ as $c^2_\alpha$, and $\sin^2\alpha$ as $s^2_\alpha$, and so forth.}
\label{tab-couplings}
\end{center}
\end{table}

\section{Flavor-Universal Analysis and Results}

\begin{table}[tb]
\begin{tabular}{ll}
\hline
Observable & Case I \\
\hline\hline
$\Gamma_Z$  & $(\Gamma_Z)_{SM}(1-0.896 c_1+0.155c_2)$ \\
$R_e$ & $(R_e)_{SM}(1+0.082c_1 + 0.912c_2)$ \\
$R_\mu$ & $(R_\mu)_{SM}(1+0.082c_1 + 0.912c_2)$ \\
$R_\tau$ & $(R_\tau)_{SM}(1+0.082c_1 + 0.912c_2)$ \\
$\sigma_h$ & $(\sigma_h)_{SM}(1-0.007c_1-0.078c_2)$ \\
$R_b$ & $(R_b)_{SM}(1-0.018c_1 - 0.201c_2)$ \\
$R_c$ & $(R_c)_{SM}(1+0.035c_1 + 0.388c_2)$ \\
$A_{FB}^e$ & $(A_{FB}^e)_{SM} + 0.176c_1 + 1.95c_2$ \\
$A_{FB}^\mu$ & $(A_{FB}^\mu)_{SM} + 0.176c_1 + 1.95c_2$ \\
$A_{FB}^\tau$ & $(A_{FB}^\tau)_{SM} + 0.176c_1 + 1.95c_2$ \\
$A_\tau(P_\tau)$ & $(A_\tau(P_\tau))_{SM} + 0.781c_1 + 8.65c_2$ \\
$A_e(P_\tau)$ & $(A_e(P_\tau))_{SM} + 0.781c_1 + 8.65c_2$ \\
$A_{FB}^b$ & $(A_{FB}^b)_{SM} + 0.555c_1 + 6.15c_2$ \\
$A_{FB}^c$ & $(A_{FB}^c)_{SM} + 0.431c_1 + 4.77c_2$ \\
$A_{LR}$ & $(A_{LR})_{SM} + 0.781c_1 + 8.65c_2$ \\
$M_W^2$ & $(M_W^2)_{SM}(1+0.430c_1 + 1.43c_2)$ \\
$M_W^2/M_Z^2$ & $(M_W^2/M_Z^2)_{SM}(1+0.430c_1 + 1.43c_2)$ \\
$g_L^2(\nu N \to \nu X)$ & $(g_L^2(\nu N \to \nu X))_{SM}+0.239(c_1 + c_2)$ \\
$g_R^2(\nu N \to \nu X)$ & $(g_R^2(\nu N \to \nu X))_{SM}-0.092(c_1 + c_2)$ \\
$g_{eV}(\nu e \to \nu e)$ & $(g_{eV}(\nu e \to \nu e))_{SM}-0.661(c_1 + c_2)$ \\
$g_{eA}(\nu e \to \nu e)$ & $(g_{eA}(\nu e \to \nu e))_{SM}$\\
$Q_W(Cs)$ & $(Q_W(Cs))_{SM}+72.7(c_1 + c_2)$ \\
\hline\hline
\end{tabular}
\caption{Corrections to precision observables in terms of the
parameters $c_1$ and $c_2$ discussed in the text.}
\label{tab-corr}
\end{table}

From the calculations in the previous two sections of the paper, 
we may compute the values of all
precisely measured electroweak quantities in terms of the Lagrangian
parameters. Using the procedure outlined
in Burgess {\it et. al.},\cite{Burgess:1993vc}  we find that the leading contributions to the observables are given in Table \ref{tab-corr}.
As mentioned earlier, all correction terms of order ${v^2\over f^2}$ are of the
forms $c_1$ and $c_2$ defined in eqn. \ref{cdefs}, meaning that they
involve the fourth power of either $\sin\omega$ or $\sin\phi$.  
In contrast, Table \ref{tab-couplings} illustrates 
the corrections of order ${v^2\over f^2}$ for Case II-IV models are suppressed by
fewer powers of these mixing angles; one expects terms 
proportional to $\sin^2\omega$ in Cases II, IV and 
to $\sin^2\phi$ in Cases III, IV.  We therefore perform a 
detailed comparison to the data only for 
models of Case I, in which all
fermions are charged under the middle two groups of the moose.

\begin{figure}[tbh]
\begin{center}
\includegraphics[width=15cm]{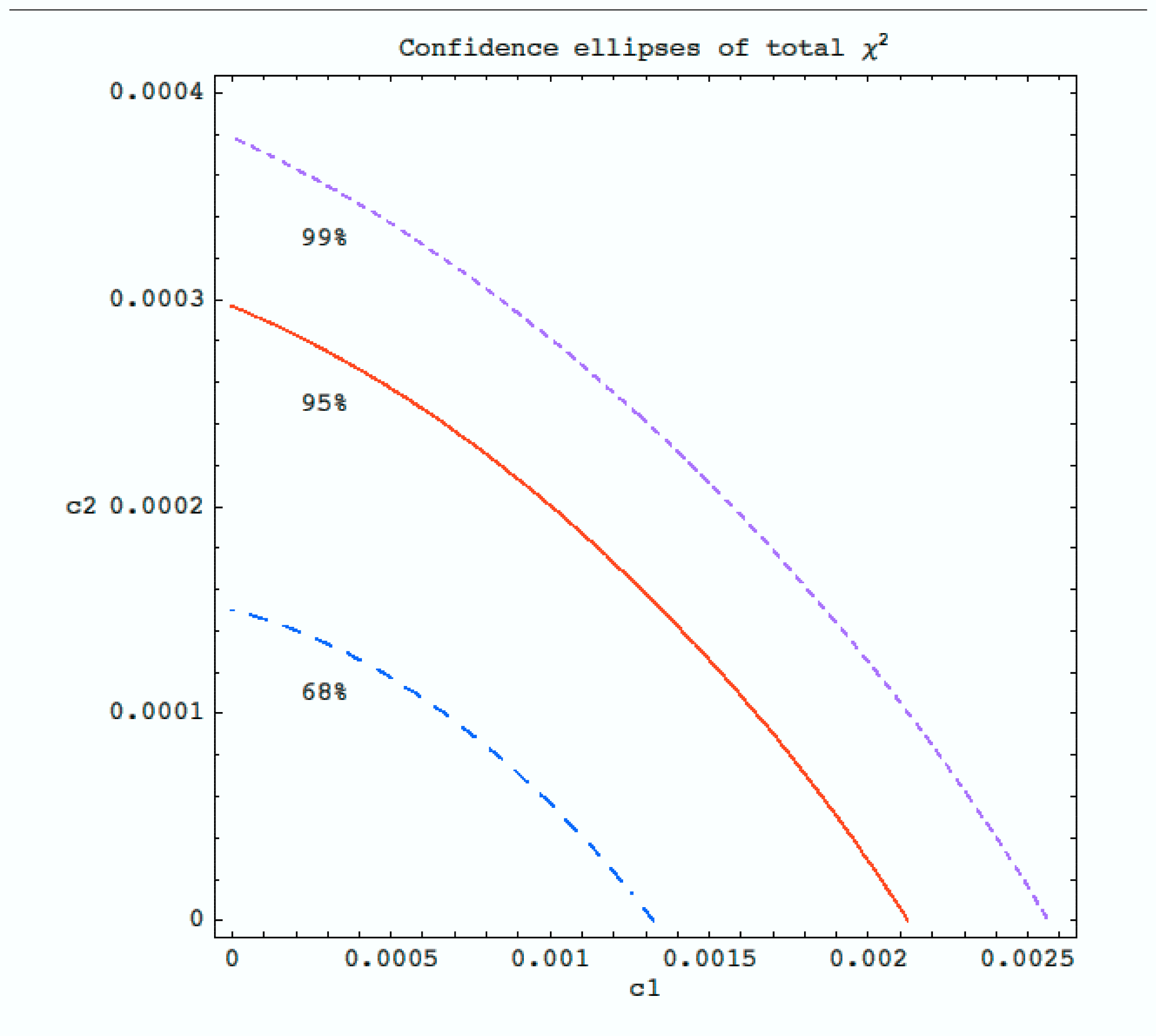}%\hspace*{8mm}
%\vspace*{-11mm}
\caption{Constraints on $c_1$ and $c_2$ at the 68\%, 95\%, and
99\% confidence level based on fits to precision electroweak 
data.\protect\cite{Group:2002mc}}
\end{center}
%\vspace*{-2.5mm}
\label{fig:Figa1}
\end{figure}

We have made a global fit of the predictions of GBESS (Case I) to 
the electroweak observables listed in the most recent
compilation by the LEP Electroweak Working 
Group \cite{Group:2002mc}.  These include Z-pole observables as well
as the width of the $W$ boson, and low-energy atomic parity violation
and neutrino-nucleon scattering. 
The 68\%, 95\%, and 99\% confidence level fits for the parameters
$c_{1,2}$ are shown in Figure 1.

The best fit values are 
\begin{equation}
c_1^{best fit} =  0.000563\ , \qquad\qquad  c_2^{best fit} =   -0.000108\ ;
\end{equation}
note that the negative value of $c_2^{best fit}$ is unphysical given the
definition \ref{cdefs} of $c_2$.   The best-fit value of $\chi^2/df$ is
30.5/20 for GBESS; the probability of a $\chi^2/df$ of this size or greater is only 6.4\%.
Fitting the same data to the standard model, the $\chi^2/df$ is 30.97/22, corresponding to a 9.7\% probability.  At present, no known model of this type gives a better fit to the data than the standard model.

Note that the constraints from the fit on $c_2$ (which depends 
on $\sin\phi$) are significantly stronger than those on $c_1$ 
(which depends on $\sin\omega$).  
In a flavor-universal model, $\sin^2\theta_{eff}$
 (the ratio of a fermion's coupling to $Q$ to its coupling to $T_3$) 
 is the same for each species.   The majority of precision electroweak
 observables can be expressed in terms of ratios of 
 the $Z$ boson's couplings to various fermions -- meaning that 
 these observables depend only on $\sin^2\theta_{eff}$.  
 Now the shift in $\sin^2\theta_{eff}$ from the standard model
 prediction (as inferred from $G_F$, $\alpha$, and $M_Z$ at tree level) depends as follows on $\sin\phi$ and $\sin\omega$
 \begin{eqnarray}
 \delta (\sin^2\theta_{eff}) \approx  &-& \left({\sin^4\theta \over \cos^2\theta - \sin^2\theta}\right) \cos^2\alpha \sin^4\omega {v^2\over f^2} \nonumber \\ &-& \left( 1 + {\sin^4\theta \over \cos^2\theta - \sin^2\theta} \right) \sin^2\alpha \sin^4\phi {v^2\over f^2} \ \ .
 \end{eqnarray}
 Since the first term in parentheses is $\sim 0.1$ and the second such term is $\sim 1.1$, then
 $\sin^2\theta_{eff}$ and the electroweak observables are far more sensitive to $\sin\phi$ than to $\sin\omega$.  

For a given value of
$\alpha$, we may unfold these constraints 
to produce a 95\% lower bound on $f$ in terms
of  $\sin\omega$ and $\sin\phi$. 
A sense of the reach of these
bounds is given in Figure 2, plotted for $\alpha=\pi/4$. For typical
values of $\sin\phi$ and $\sin\omega$, the bounds on the scale $f$ range from a few TeV. 
Note that the standard model structure is formally recovered in the limit where
both $\sin\phi$ and $\sin\omega$ approach zero.

\section{Discussion and Conclusions}

In studying the Generalized BESS class of models with extended electroweak
gauge groups, we have seen how the assignment of fermion charges affects the structure of corrections to precision electroweak observables.   Having fermions transform under the outer $SU(2)$ or $U(1)$ group of the linear moose (as in Cases II-IV) results in leading corrections that are suppressed by only two powers of the sine of the weak ($\omega$) or hypercharge ($\phi$) mixing angle, respectively.  If all fermions are charged only under the innermost groups (as in Case I), all corrections to  electroweak observables are instead suppressed by $\sin^4 \omega$ or $\sin^4 \phi$. 

This fourth-power suppression was previously noted in the so-called {\it degenerate BESS} model \cite{Casalbuoni:1995qt}
in which a parity symmetry imposed the condition $\alpha=\pi/4$.   Our
results show that the suppression is independent of $\alpha$, and
arises purely from the choice of gauge charges for the fermions.  Degenerate BESS exhibits the suppression not because of the parity symmetry (as originally supposed) but because its fermion couplings are as in Case I.

As noted earlier, a number of models in the literature\cite{Chivukula:1994mn,Chivukula:1995gu,Georgi:1989ic,Georgi:1989xz,Hill:1991at,Hill:1994hp,Pisano:ee,Frampton:1992wt,Dimopoulos:2002mv} are special cases of the GBESS framework.  We can now understand the structure of the electroweak corrections found in the literature on these models in terms of the general properties outlined above.  Many of these models are flavor non-universal, in contrast to the universal case discussed here.   In such cases, one applies the rules outlined here separately to each species of fermion.  For example, consider the heavy case of NCETC (equation \ref{eq:ncetcfig}).  This is an example of GBESS in the limit where $(\sin\alpha,\, \sin\phi) \to 0$ because there is no fourth group in the moose.   The first and second generation fermions behave as in Case II of GBESS because they couple to the outer $SU(2)_2$ group; the 3rd generation fermions behave as in Case I.  Therefore we predict the Fermi constant to be $\sqrt{2} G_F = v^{-2} + f^{-2}$ and the $Z$ coupling to left-handed fermions to be
\begin{equation}
{e\over \sin\theta \cos\theta} \left[ 1 + {v^2\over f^2}(\cos^2\omega\, \sin^2\omega\, T_3^L - \sin^4\omega\, T_3^H)\right]
\end{equation}  
where the separate weak generators for the heavy (H) and light (F) fermions are distinguished.
Both of these results  match what was found in \cite{Chivukula:1995gu}.   In the light case of NCETC, it is the light fermions which behave as in our Case I and the heavy fermions as in Case II; the predictions for $G_F$ and the $Z$ boson coupling to fermions are modified accordingly.\footnote{In making a comparison with light NCETC \cite{Chivukula:1995gu}, one must note the following difference of conventions.  
In our discussion, we move fermions from one gauge group to another, leaving the order of the moose and the names of the couplings unaltered.  In changing from heavy to light NCETC the original reference effectively moved the $U(1)$ hypercharge group to the left-hand end of the moose and swapped the couplings $g$ and $\tilde g$.  Thus to  compare with this reference we must make the exchange $\sin\omega \to \cos\omega$.}

Many of the specific models cited above correspond to taking a weak-coupling limit for
the extra gauge groups (i.e., taking  $\sin\phi$ or $\sin\omega$ of order 1),
in which case the bounds on $f$ are of order 10 TeV, as consistent with results in the literature for topcolor, NCETC, UUM and electroweak $SU(3)$ \cite{Chivukula:1995gu,Chivukula:1996cc,Chivukula:2002ry,Csaki:2002bz}.  This tends to make the extra gauge bosons too heavy to observe easily in collider experiments.  
Formally the
corrections  to electroweak observables vanish when the couplings of the extra gauge groups
become strong, that is in the limit $(\sin\phi\,, \sin\omega) \to 0$. 
The phenomenologically interesting question is whether there
are {\it any} viable models corresponding to this limit, which would result in interesting structure at accessible scales.

\begin{figure}[tbh]
\label{fig:Fig2}
\begin{center}
\includegraphics[width=15cm]{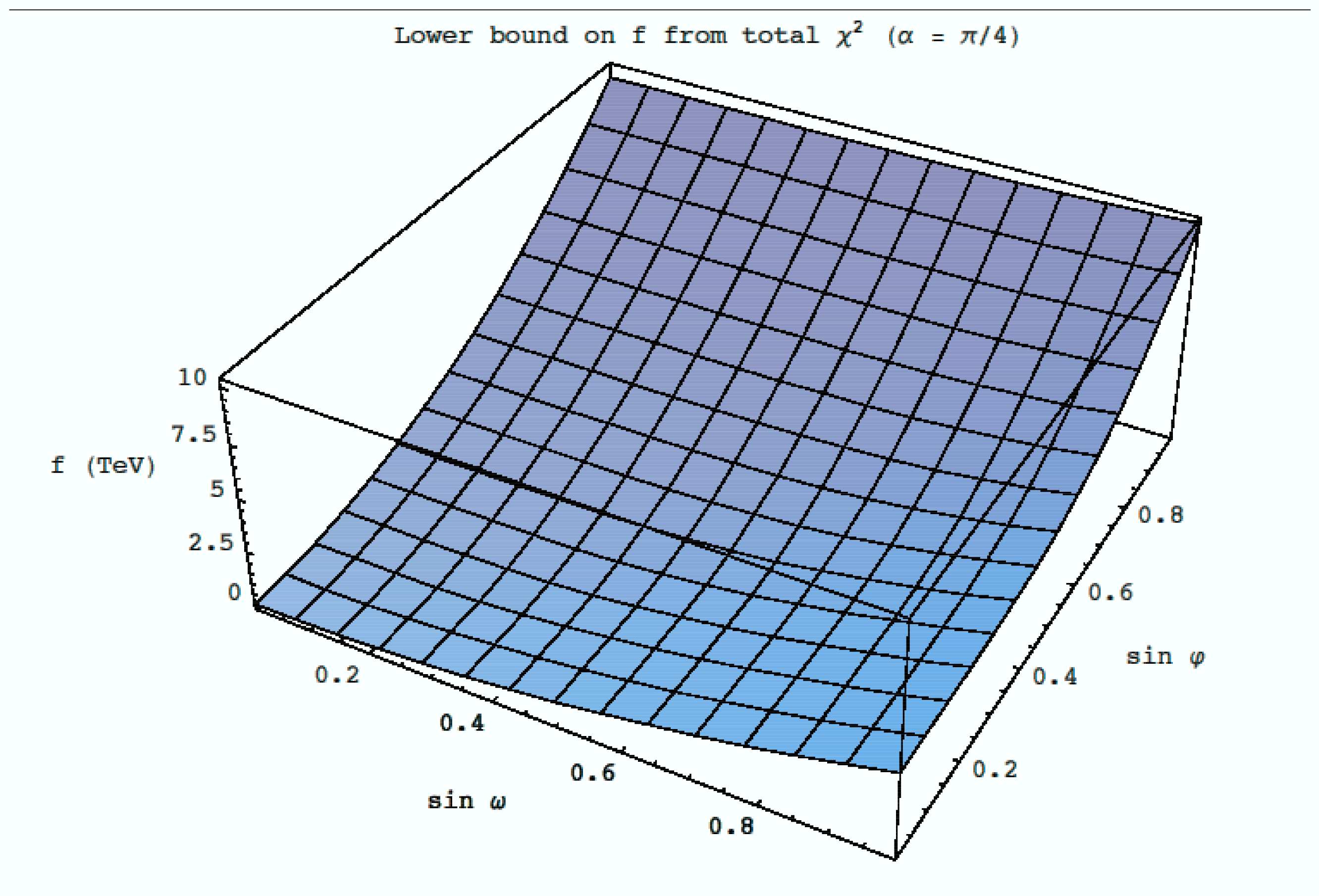}%\hspace*{8mm}
%\vspace*{-11mm}
\caption{
The lower-bound on $f$
at 95\% confidence level for $\alpha=\pi/4$, as a function of $\sin\omega$
and $\sin\phi$,  based on fits to precision electroweak 
data.\protect\cite{Group:2002mc}
}
\end{center}
%\vspace*{-2.5mm}
\end{figure}

\section*{Acknowledgments}
\uppercase{W}ork partially
supported by the \uppercase{US} \uppercase{D}epartment of \uppercase{E}nergy
under grants \uppercase{DE-FG02-91ER40676} and \uppercase{DE-FG03-93ER40757}.
E.H.S. and R.S.C. acknowledge the hospitality of the Aspen Center for Physics where part of this work was completed.

\end{document}